\documentclass{aa}  

\usepackage{graphicx}
\usepackage{natbib}
\usepackage[varg]{txfonts}
\usepackage{caption}
\usepackage{subcaption}
\usepackage{appendix}
\usepackage{rotating}
\usepackage{longtable}
\usepackage{lscape}
\usepackage{footnote}

\begin{document}

   \title{Laboratory oscillator strengths of \ion{Sc}{I} in the near-infrared region for astrophysical applications}

   \author{A. Pehlivan
          \inst{1,2}
          \and
          H. Nilsson\inst{2}
          \and
          H. Hartman\inst{1,2}
          }

   \institute{Material Science and Applied Mathematics, Malm\"{o} University, SE-205 06 Malm\"{o}, Sweden \\
             \email{asli.pehlivan@mah.se, asli@astro.lu.se} 
   \and
         Lund Observatory, Box 43, SE-221 00 Lund, Sweden\\
             }

   \date{Received 23 June 2015 / Accepted 25 August 2015 }

  \abstract
   {Atomic data is crucial for astrophysical investigations. To understand the formation and evolution of stars, we need to analyse their observed spectra. Analysing a spectrum of a star requires information about the properties of atomic lines, such as wavelengths and oscillator strengths. However, atomic data of some elements are scarce, particularly in the infrared region, and this paper is part of an effort to improve the situation on near-IR atomic data.}
   {This paper investigates the spectrum of neutral scandium, \ion{Sc}{I}, from laboratory measurements and improves the atomic data of \ion{Sc}{I} lines in the infrared region covering lines in R, I, J, and K bands. Especially, we focus on measuring oscillator strengths for \ion{Sc}{I} lines connecting the levels with 4p and 4s configurations.}
   {We combined experimental branching fractions with radiative lifetimes from the literature to derive oscillator strengths (\textit{f}-values). Intensity-calibrated spectra with high spectral resolution were recorded with Fourier transform spectrometer from a hollow cathode discharge lamp. The spectra were used to derive accurate oscillator strengths  and wavelengths for \ion{Sc}{I} lines, with emphasis on the infrared region. }
   {This project provides the first set of experimental \ion{Sc}{I} lines in the near-infrared region for accurate spectral analysis of astronomical objects. We derived 63 log$(gf)$ values for the lines between $5300{\AA}$ and $24300{\AA}$. The uncertainties in the \textit{f}-values vary from $5\%$ to $20\%$. The small uncertainties in our values allow for an increased accuracy in astrophysical abundance determinations.}
   {}


   \maketitle
%

\section{Introduction}

Knowing the Milky Way's chemical composition and its chemical abundance can lead us to understand how the Milky Way has formed and evolved. With time, the amount of heavy elements in a galaxy increases as new stars are born and die. Studies of elemental composition of stars will give information about the environment in which they were born. This, in turn, helps us to construct the evolution history of our Galaxy. For this reason, it is important to derive accurate abundances of stars.\\
\indent Elemental abundances can be derived by comparing an observed stellar spectrum with a synthetic spectrum. This approach requires atomic data, meaning oscillator strengths (\textit{f}-values) and wavelengths. The measured equivalent width of a line is directly proportional to the abundance of the element and the \textit{f}-value. The lack of atomic data makes it difficult to compare the observed and synthetic stellar spectra. Thus, an accurate and complete set of oscillator strengths and wavelengths are crucial for abundance analysis.\\
\indent
In the periodic table, scandium is between the $\alpha$-elements, e.g. O, Mg, Si, S, Ca, and the iron peak elements, such as V, Cr, Mn, Fe. The formation scenarios of $\alpha$-elements and iron peak elements have been studied in detail. Stars that end their lives in supernova type II explosions are responsible for producing $\alpha$-elements, whereas stars that die in supernova type Ia contribute to the formation of iron peak elements and yet there is no clear picture of how scandium is formed \citep{Nissen,Francois,Zhang,Chiara}. \\
\indent
Studies of F and G stars may provide insight into the creation of scandium. However, the abundance varies in these stars.  There seems to be a disagreement in the scandium abundance pattern. In some studies, scandium is overabundant relative to iron \citep{Zhao, Nissen}, whereas the others found no deviations \citep{Gratton, Prochaska}. Furthermore, scandium is an important element for understanding Am and Fm stars, which are overabundant in iron peak elements but underabundant in Sc and Ca \citep{Leblanc,Alecian}. In addition to these, some studies cannot be completed owing to the quality of oscillator strengths in the infrared region \citep{Schmidt}. Improved atomic data with small uncertainties in oscillator strengths can help in solving these problems.\\
\indent
The scandium abundance in the Sun was determined by \citet{Grevesse} to be $3.18\pm 0.10$ dex, and by \citet{Zhang} $3.13\pm0.05$ dex and by \citet{Asplund} to be $3.15\pm0.04$ dex. These values are consistent within the error bars. A high quality laboratory measurement of oscillator strengths with small uncertainties is a requirement for determining the accurate abundance of scandium in the solar photosphere.\\
\indent
In astrophysical applications, log$(gf)$ values are usually used, where $g$ is the statistical weight of the lower level and $f$ is the oscillator strength (\textit{f}-value). \citet{Parkinson} derived the experimental log$(gf)$ values for \ion{Sc}{I} lines involving the ground state in the optical region.  \citet{Lawler} measured experimental log$(gf)$ values for lines in the optical region involving the ground state or to the $4$s $^4$F term. For other transitions, generally semi-empirical values of \citet{Kurucz} are the favoured ones. In addition, \cite{Ozturk} calculated the oscillator strengths of neutral scandium with the Cowan code and the quantum defect orbital code (CDOT) methods but did not include any near-infrared transitions. To our knowledge, there are no experimental oscillator strengths of \ion{Sc}{I} lines in the infrared region.\\
\indent
In recent years, infrared spectroscopy has become important for observing astronomical objects. Both new generation of ground-based telescopes and space telescopes, which work in the  infrared region, demand complete and accurate atomic data. Atomic data of scandium are scarce in the infrared region and missing data constrain analysis of a stellar spectrum recorded with new-generation advanced instruments and, as a result, limit the construction of a galactic evolution model. \\
\indent
In this work, we measured accurate wavelengths and oscillator strengths of \ion{Sc}{I} lines from high-resolution laboratory measurements in the infrared region from the upper odd parity $4$p $^4$F$^o$, $^4$D$^o$, $^2$D$^o$, $^2$P$^o$, $^4$G$^o$, $^2$G$^o$ and even parity $4$s $^2$P, $^2$S terms.
Section 2 describes the branching fraction measurements. In addition, this section includes the experimental setup we used during our measurements, calibration of wavenumbers, and the uncertainty calculations. In section 3 we present our results with their uncertainties and a comparison of our results with previous studies.\\


\section{Materials and methods}

\subsection{Branching fraction measurements}

The oscillator strength, \textit{f-value}, is related to the transition probability for electric dipole transitions by
\begin{equation}\label{eq:eq1}
f=\frac{g_u}{g_l} \lambda^2 A_{ul} 1.499 \hspace{0.5mm} \cdot \hspace{0.5mm} 10^{-16},
\end{equation}
where $g_u$ is the statistical weight of the upper level, $g_l$ the statistical weight of the lower level, $\lambda$ the wavelength of the transition in \AA, and $A_{ul}$ the transition probability in s$^{-1}$ between the upper level $u$  and the lower level $l$. \\
\indent The branching fraction (\textit{BF}) for a given transition from the upper level $u$ to the lower level $l$ is defined as the ratio of the transition probability, $A_{ul}$, of the transition to the sum of all transitions from the same upper level to all lower states $i$. The transition probability is proportional to the transition line intensity; therefore,  
\begin{equation}\label{eq:eq2}
BF_{ul}= \frac{A_{ul}}{\sum_{\rm i} A_{ui}}= \frac{I_{ul}}{\sum_{\rm i} I_{ui}}.
\end{equation}
\indent Lifetime measurements are complementary to \textit{BF} derivations. The radiative lifetime of the upper level, $\tau_u$,  is the inverse sum of all transition probabilities from the same upper level $\tau_u= 1/\sum_{\rm i} A_{ui}.$
The transition probability of a line can be derived by combining experimental \textit{BF}s and radiative lifetimes;
\begin{equation}\label{eq:eq4}
A_{ul}= \frac{BF_{ul}}{\tau_u}.
\end{equation}
We derived transition probabilities, $A_{ul}$, with the help of (Eq. 3) and converted these values to \textit{f}-values using (Eq. 1) .\\
\indent To derive accurate \textit{BF}s, all transitions from the same upper level should be included. In this work, some upper energy levels had transitions in two different spectral regions recorded with different detectors. In such cases, because \textit{BF}s require only the relative intensities, the spectra have to be put on a same relative intensity scale by using a normalisation factor. This normalisation factor was determined from the \ion{Sc}{I} lines, which were visible in both spectra. \\
\indent Fig. 1 shows the partial energy level diagram of \ion{Sc}{I} levels. In this figure we have marked the transitions that appeared in our spectra. Transition lines from the same upper level were predicted from the \citet{Kurucz} database and from \citet{Benahmed}. These lines were identified in our spectra by using FTS analysis software GFit \citep{Engstrom,Engstrom2}. 

        \begin{figure*}
   \centering
\includegraphics[width=\textwidth]{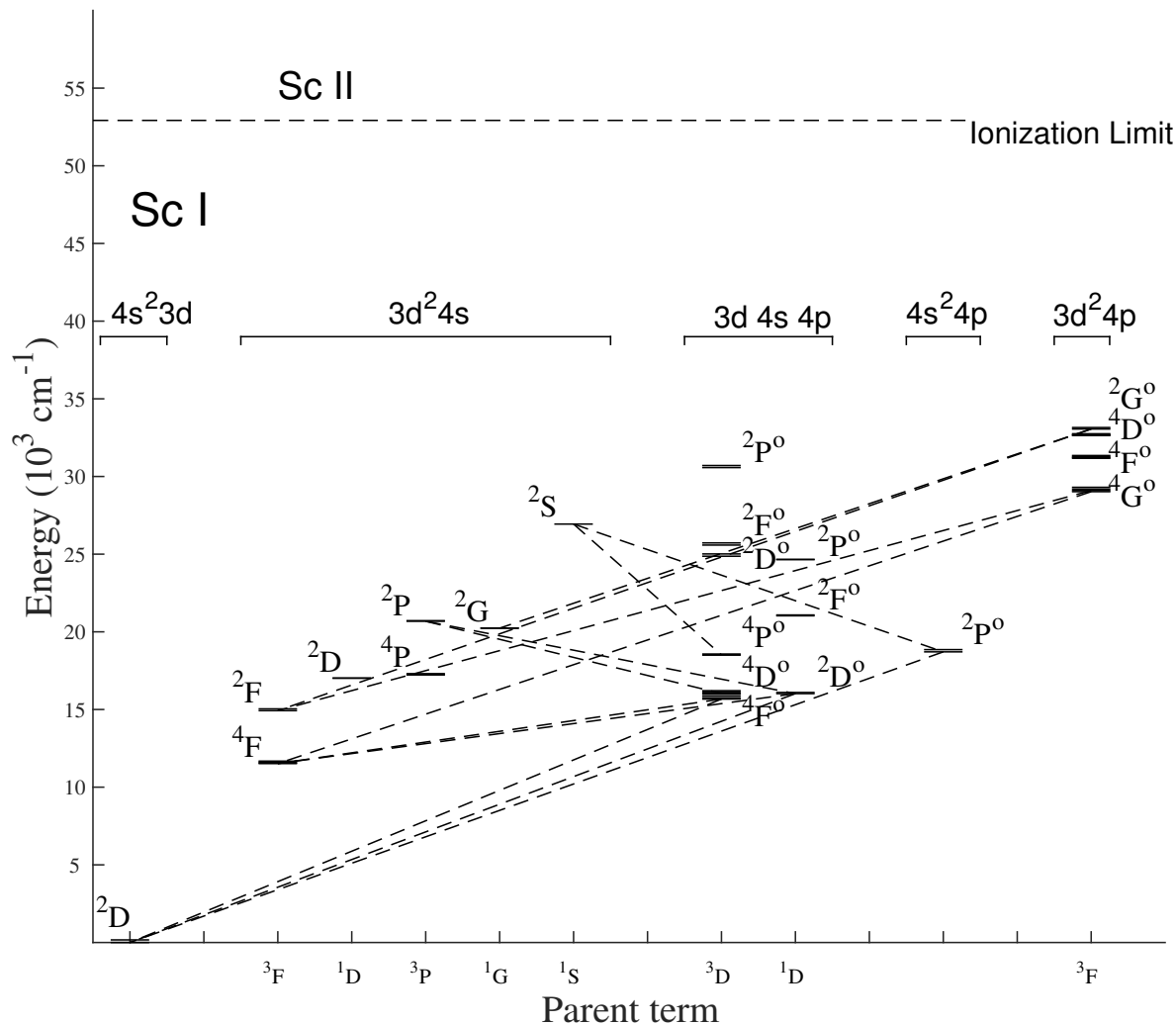}
  \caption{Partial energy level diagram of \ion{Sc}{I} levels showing the observed transitions in the measurements.}
              \label{FigGam}
    \end{figure*}
       
Sc I has only one stable isotope, \element[ ][45]{Sc}, with a nuclear spin I$=7/2$, so \ion{Sc}{I} lines show hyperfine splitting (hfs). Interaction between the nuclear spin and the total angular momentum splits energy levels into the hyperfine structure. This splitting appears as a resolved line splitting or a line broadening in a spectrum. Unsuccessful treatment of hfs leads to incorrect abundances, as much as several tenths of a dex \citep{Nielsen}. For lines with negligible hfs, one can fit a Voigt profile to the line and measure the central wavenumber and intensity. However, this is not the case for hfs. Having hfs gives rise to a difficulty in fitting a line profile since every fine-structure transition appears as a broadened or multi-component feature. For this reason, we defined the position of lines by the centre of gravity and instead of fitting a Voigt profile, we measured the integrated intensity under each measured line with the help of GFit. We applied this method for the cases in which we have a single line profile i.e. to test the accuracy. In all of these cases, fitting a line profile and taking an integrated intensity gave the same result.  \\

\subsection{Experimental setup}

A water-cooled hollow cathode discharge lamp (HCL) was used as a source to produce scandium atoms. The HCL consists of a glass tube, anodes on each side, and an iron cathode in the middle. A small solid sample of scandium was placed in the iron cathode. The diameter of the inner cathode was 7 mm and the distance between anode and cathode was 20 mm. The light source was run with a pressure of 1.0 Torr (1.3 mbar), with different applied currents, ranges from 0.10 A to 0.60 A, and with argon or neon as a carrier gas. The best condition, i.e. the strongest lines, for the measurements was achieved with argon gas and with applied current of 0.60 A. The argon lines were used as reference lines for wavenumber calibration as described in section 2.3.\\
\begin{figure}
        \centering
                \includegraphics[width=0.5\textwidth]{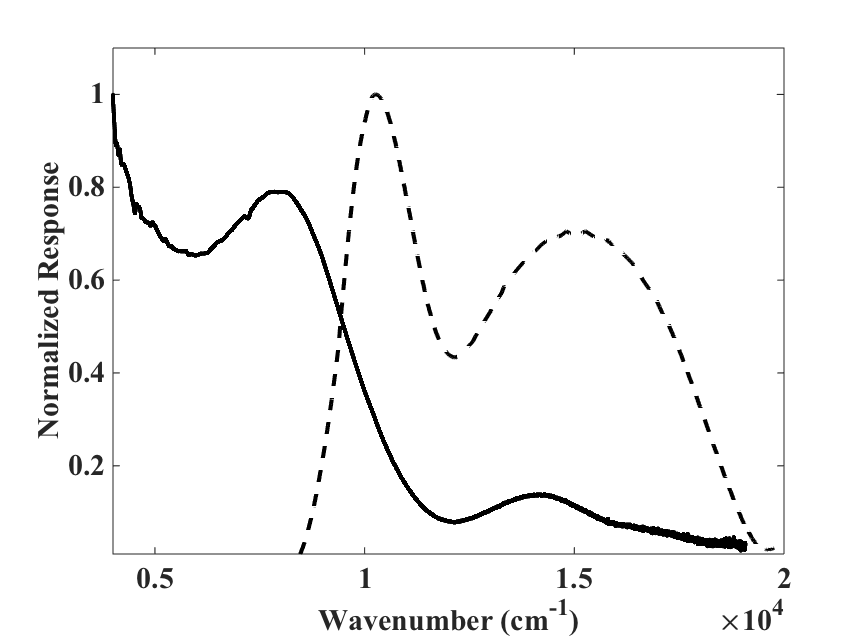}
                \caption{Instrument response function with InSb (solid line) and Si (dashed line) detectors, respectively. Each curve is normalised to its highest point. }
                \label{fig:figa}
                  \end{figure}
\indent
The \ion{Sc}{I} spectra were recorded using the high resolution Fourier transform spectrometer (FTS), Bruker IFS 125 HR, at the Lund Observatory (Edl{\'e}n Laboratory). This instrument has the wavenumber range of $50000-2000$ cm$^{-1}$ ($200-5000$ nm) and the maximum resolving power of $10^6$ at $2000$ cm$^{-1}$. The resolution of the FTS was set to $0.02$ cm$^{-1}$ during measurements. We recorded the spectra with the indium antimonide (InSb) and silicon (Si) detectors, which are sensitive to different spectral regions.\\
\indent The optical elements in FTS contribute to the response function of the instrument. Observed line intensities differ from the intrinsic line intensities, due to the wavelength dependent transmission of the spectrometer and optical elements. It was therefore required to find the response functions of two different detectors. After determining the response function, one can acquire the intrinsic line intensities by dividing the observed line intensities with the response function. A common way to determine the response function is to use an intensity calibrated reference. We used a tungsten filament lamp to calibrate the intensities of \ion{Sc}{I} lines. This lamp in turn has been calibrated by the Swedish National Laboratory (SP) for spectral radiance in the range $40000-4000$ cm$^{-1}$ ($250-2500$ nm). If the radiance of the reference is known, one can determine the response function. Fig. 2 shows the response function of the instrument with Si and with InSb detectors. \\
\indent
We recorded the spectrum of the tungsten lamp immediately before and after every scandium recording to verify the response of the instrument did not change during the measurements. For this reason the HCL and tungsten lamp were placed at the same distance from the FTS and a rotating mirror was used in order to change the light source (HCL or the Tungsten lamp) without moving the lamps. There were no changes in the spectra of the tungsten lamp taken before and after the HCL measurements, therefore one response function for each detector was used for calibration. Using intensity calibrated spectra, relative line intensities at the different regions were connected with a normalisation factor. This was done by using the \ion{Sc}{I} lines in the overlapping region of the two detectors, see Fig. 2.\\
\indent
The spectra of \ion{Sc}{I} lines were recorded with different currents. Higher current values help weaker lines to be visible, but introduce self-absorption. Self-absorption has a negative effect on line intensities e.g. it changes the observed line profile and gives an incorrect intensity. This in turn affects the \textit{BF} measurements, see Eq. \eqref{eq:eq2}.  The effect of self-absorption depends on the number density of atoms in the plasma and intensity of the line. The number density of the plasma increases by increasing the current applied to the HCL. The higher concentration of scandium ions in the plasma results in more lines affected by self-absorption. We performed a set of measurements with varying currents to determine if \ion{Sc}{I} line intensities were affected by self-absorption at higher currents. Then we plotted normalised integrated intensity ratios of lines from the same upper level versus applied current. Integrated intensity ratios between lines from the same upper level must be same for different applied currents if there is no self-absorption. In our measurements some lines showed effects of self-absorption. In these cases, we extrapolated integrated intensity ratios to zero current, where self-absorption is assumed to be zero, and used these values to obtain correct relative intensities. \citet{Sikstrom} tested this method by comparing the self absorbed Fe I lines intensity ratios at zero current with the ratios derived by \citet{Blackwell}  using absorption measurement. They found that the extrapolated intensity ratios agree with the \citet{Blackwell} intensity ratios. The correction of self-absorption introduce uncertainties in the \textit{BF} determination.\\

\subsection{Wavenumber calibration}

The HeNe laser in the FTS determines the displacement of the FTS moving mirror and gives a multiplicative wavenumber scale to the spectrum. However, a non-parallel alignment with the incoming light beam adds a shift to the measured wavenumber. In addition to this difference, having a finite size aperture may lead to wavenumber shifts \citep{Learner}. The shift introduced by these effects is multiplicative. With a correction factor this shift can be calibrated;
\begin{equation}\label{eq:eq5}
\sigma_{true} = \sigma_{apparent}(1+k_{eff}).
\end{equation}
In Eq. \eqref{eq:eq5},  $\sigma_{true}$ denotes the corrected wavenumber, $\sigma_{apparent}$ is the measured wavenumber, and $k_{eff}$ is the correction factor. We used the wavenumbers of \ion{Ar}{i} lines from \citet{Whaling}, including the multiplicative correction by \citet{Sansonetti}, as a wavenumber standard for the wavenumber calibration of \ion{Sc}{I} lines. \\
\indent In principle, using only one calibration line is enough to fix the multiplicative constant, but several calibration lines can be employed for the purpose of increasing the wavenumber accuracy. We calculated a correction factor for each \ion{Ar}{i} line and used the weighted mean of these factors. \\ 
\indent We combined the errors in the determination of the multiplicative factor and the errors from centroiding the position of each line to estimate an absolute uncertainty of the observed wavenumbers of $u(\sigma)$=$\pm0.001$ cm$^{-1}$.

\subsection{Uncertainties}

The contribution to the \textit{f-value} uncertainties comes from the uncertainty of the radiative lifetime and the uncertainty of the branching fractions. The branching fraction uncertainties contain the uncertainty of the intensity calibration lamp and the uncertainty that arises from using two different detectors, i.e. uncertainty of the normalisation factor and the contribution from the self-absorption correction. \\
\indent The uncertainty of the line intensity (or the integrated intensity) and the uncertainty of the radiative lifetime are uncorrelated. Although \textit{BF}s are dependent on each other, we treated them as independent in the uncertainty calculations. By including all the uncertainties from different effects described above, one can determine the total uncertainty of the \textit{BF}s as specified in \citet{Sikstrom},
\begin{align} 
\Bigg(\frac{u(BF)}{BF} \Bigg)^2 &= (1-(BF)_k)^2 \hspace{0.6mm}\Bigg(\frac{u(I_k)}{I_k} \Bigg)^2  \nonumber \\
&+ \sum\limits_{j \neq k(inP)} (BF)_j^2 \hspace{0.8mm} \Bigg( \bigg( \frac {u(I_j)}{I_j} \bigg )^2 + \bigg( \frac{u(c_j)}{c_j} \bigg)^2 \Bigg) \nonumber \\
&+ \sum\limits_{j \neq k(inQ)} (BF)_j^2 \hspace{0.8mm} \Bigg( \bigg( \frac {u(I_j)}{I_j} \bigg )^2 + \bigg( \frac{u(c_j)}{c_j} \bigg)^2 +\bigg(\frac{u(nf)}{nf} \bigg)^2 \Bigg).
\end{align}
In the first term on the right-hand side of Eq. 7,  $BF_k$ denotes the branching fraction of the line in question in the spectral region of the detector $P$ and $u(I_k)$ denotes the uncertainty in the measured intensity of the line k. The second term includes the \textit{BF}s and measured intensity uncertainties of the other lines from the same upper level in the spectral region of detector $P$ and the uncertainty of the calibration lamp $u(c_j)$. The last term contains the \textit{BF}s and measured intensity uncertainties of the lines in the spectral region of detector $Q$, the uncertainty of the calibration lamp, and the uncertainty of the normalisation factor, $u(nf)$, between the spectral regions. The uncertainty of the calibration lamp, $u(c_j)$, is $7\%$ and the uncertainty of the normalisation factor is $u(nf)= 5\%$. We used the uncertainties of the integrated intensity, $u(I)$, from GFit and the relative uncertainty varies between $0.002\%$ for the strong lines and $10\%$ for the two weak lines and self-absorbed lines. The cases in which we had self absorbed lines, we included an uncertainty from self absorption correction as well. This uncertainty varies between $1\%$ and $9\%$. From Eq. 3, the uncertainty of the \textit{f}-values or the transition probability becomes
\begin{equation}
\Bigg(\frac{u(f_k)}{f_k} \Bigg)^2=\Bigg(\frac{u(A_k)}{A_k} \Bigg)^2= \Bigg(\frac{u(BF)}{BF} \Bigg)^2 + \Bigg(\frac{u(\tau)}{\tau} \Bigg )^2,
\end{equation}
where $u(\tau)$ is the uncertainty of the radiative lifetime of the upper level. The derived uncertainties of \textit{f}-values range between 5$\%$ for the strong lines and $20\%$ for the weak lines or for the lines with high uncertainty in their radiative lifetime. \\
\indent Two different sets of radiative lifetimes were used, one set of experimental lifetimes and one set of semi-empirical lifetimes. The uncertainty of the experimental lifetimes is $5\%$ \citep{Marsden}. For semi-empirical values, we compared the semi-empirical lifetimes \citep{Kurucz} of the levels that have experimental lifetime values and examined how they differ from the experimental lifetime values. Based on this comparison we adopted $20\%$ relative uncertainty for the semi-empirical lifetimes. 
\begin{figure}
        \centering
        \begin{subfigure}{0.52\textwidth}
                \includegraphics[width=\textwidth]{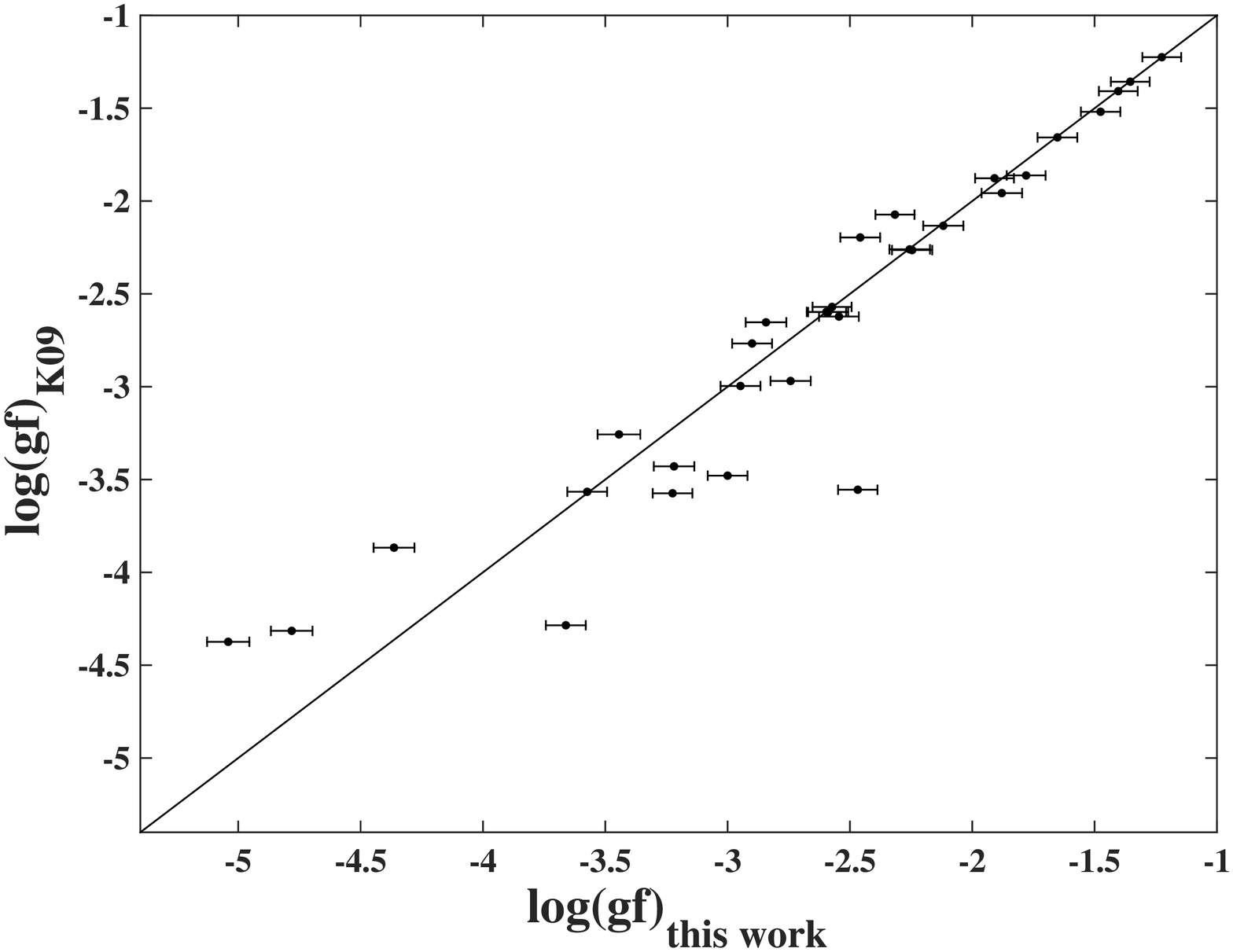}
                \caption{with semi-empirical radiative lifetimes}
                \label{fig:comp1a}
        \end{subfigure}
        \begin{subfigure}{0.52\textwidth}
                \includegraphics[width=\textwidth]{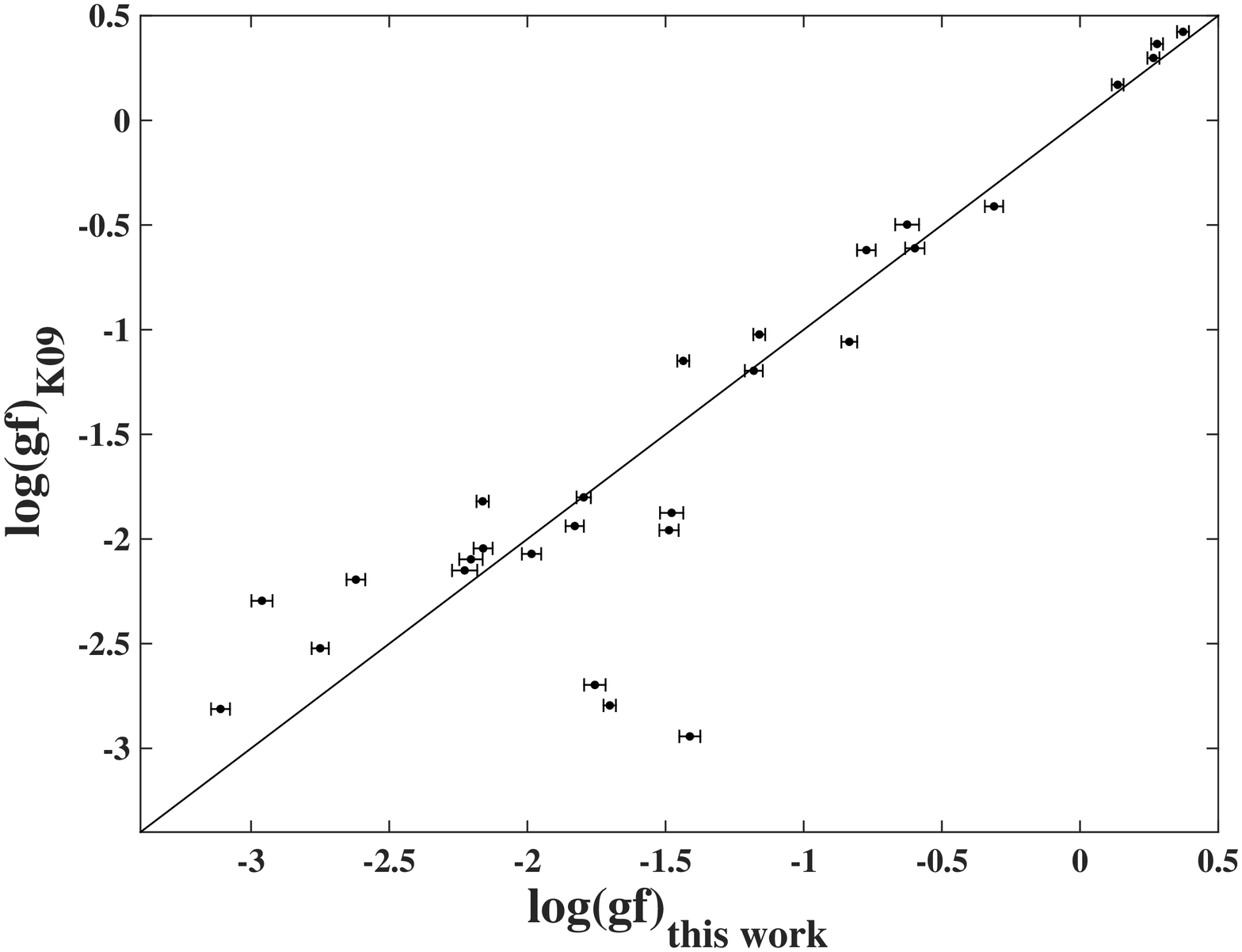}
                \caption{with experimental radiative lifetimes}
                \label{fig:comp1b}
        \end{subfigure}
     \caption{Comparison between log$(gf)$ values of this work and the semi-empirical \citet{Kurucz} log$(gf)_{K09}$ values. Upper panel (a) includes the lines with semi-empirical lifetimes, whereas the lower panel (b) includes lines with experimental lifetimes.}
\end{figure}

\begin{figure}
        \centering
        \begin{subfigure}{0.52\textwidth}
                \includegraphics[width=\textwidth]{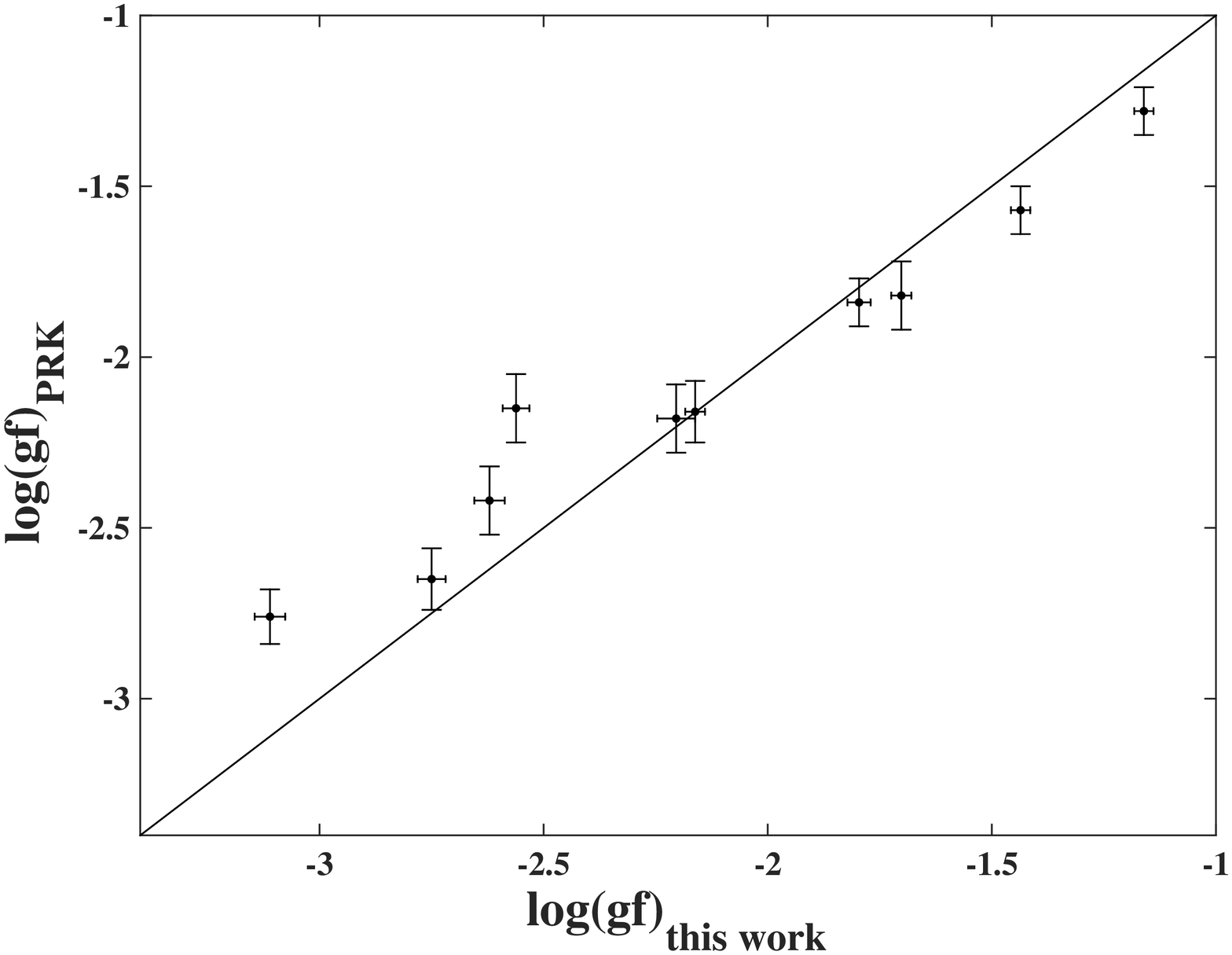}
                \caption{with \citet{Parkinson} log$(gf)$ values}
                \label{fig:comp2a}
        \end{subfigure}
        \begin{subfigure}{0.52\textwidth}
                \includegraphics[width=\textwidth]{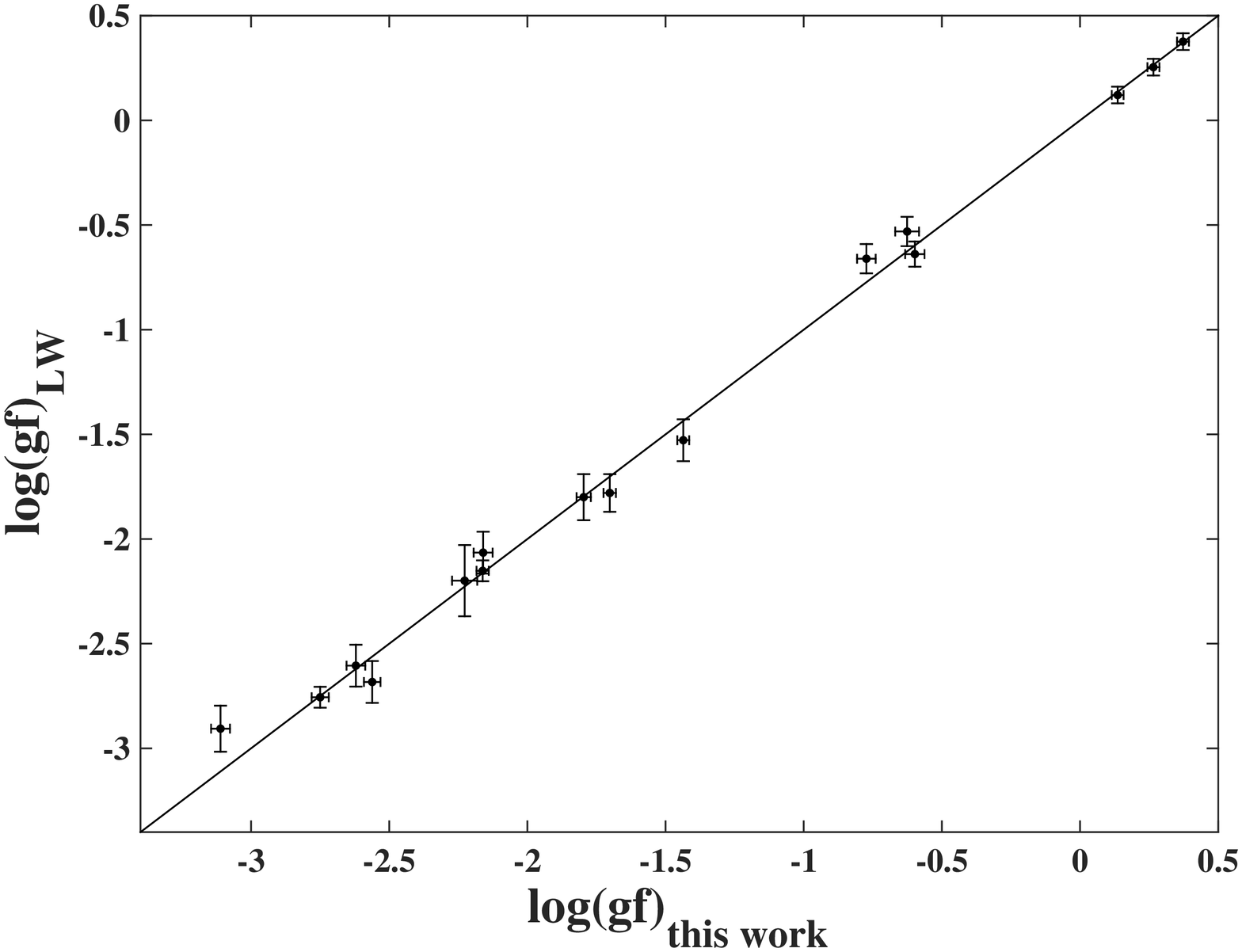}
                \caption{with \citet{Lawler} log$(gf)$ values}
                \label{fig:comp2b}
        \end{subfigure}
     \caption{Comparison between log$(gf)$ values of this work derived from the experimental radiative lifetimes and the previous experimental log$(gf)_{PRK}$ values of \citet{Parkinson} and log$(gf)_{LW}$ values of \citet{Lawler}, in subplot (a) and (b), respectively. These studies do not include infrared transitions.}
\end{figure}


\section{Results and conclusion}

We calculated the \textit{BF}s from observed line intensities from Eq. 2. All the lines from the same upper level should be included to have accurate \textit{BF}s. For this purpose we recorded several spectra with different detectors, which are sensitive to different regions. We used the lines in the overlapping regions of two different detectors to put lines on the same relative intensity scale.  Some of the weak lines were not observed in the spectra. In these cases we used theoretical transition probabilities to estimate the residual value, i.e. the missing \textit{BF} from the unobserved lines, for these levels. The values are less than one per cent for all levels, except from the 4s $^2$S$_{1/2}$ level. Derived log$(gf)$ values for the lines together with their uncertainties and the residual values of missing lines are given in Table 1 and Table 2. \\
\indent In addition to using different detectors, the HCL was run with different discharge currents. This helped us to find out if the strong lines were affected by self-absorption. In the cases in which there were lines affected by self-absorption, we corrected the measured intensities of these lines. This in turn added an extra uncertainty to the \textit{BF} of the affected line. \\
\indent We derived transition probabilities, $A_{ul}$, by combining \textit{BF}s and radiative lifetimes. Where available the radiative lifetimes are taken from the measurements of \citet{Marsden} and the rest of the lifetimes are semi-empirical values from \citet{Kurucz}. The dominating factor in the uncertainty of oscillator strengths comes from the radiative lifetimes, resulting in higher uncertainties for the oscillator strengths derived from the semi-empirical lifetimes. They could be improved with laboratory measurements.\\
\indent We derived oscillator strengths for 63 lines from $4$p $^4$F$^o$$_{3/2,5/2,7/2,9/2}$, $^4$D$^o$$_{1/2,3/2,5/2,7/2}$, $^2$D$^o$$_{{3/2},5/2}$, $^2$P$^o$$_{3/2}$, $^4$G$^o$$_{5/2,7/2,9/2}$, $^2$G$^o$$_{9/2}$, and $4$s $^2$P$_{1/2,3/2}$, $^2$S$_{1/2}$ levels with uncertainties in the \textit{f}-values between 5$\%$ for the strong lines and $20\%$ for the weak lines or for the lines with high uncertainty in their radiative lifetime. These results together with their branching fractions, transition probabilities, and previously published log$(gf)$ values are presented in Table 1 and Table 2. The first column of the previously published log$(gf)$ values is the semi-empirical calculations of \citet{Kurucz} and the other two columns are experimental values of \citet{Parkinson} and \citet{Lawler}. These previous measurements only included lines for the optical region and to our knowledge this is the first time that the experimental log$(gf)$ values of \ion{Sc}{I} lines in the infrared region were measured. Table 1 shows our results derived from the experimental radiative lifetimes and Table 2 from the semi-empirical radiative lifetimes.\\
\indent In Figs. 3 and 4, we graphically compare our data with previously published studies. Fig. 3 shows our results compared with \citet{Kurucz} semi-empirical log$(gf)_{K09}$ values and Fig. 4 presents our results compared to experimental log$(gf)_{PRK}$ of \citet{Parkinson} and log$(gf)_{LW}$ \citet{Lawler}. The log$(gf)$ values in Fig. 3a were derived from the semi-empirical radiative lifetimes, and in Fig. 3b they were derived from the experimental values. As seen in Figs. 3a and 3b, the log$(gf)$ values measured in this work agree with the semi-empirical values for high log$(gf)$ values. For lower values the scatter is larger. This can be explained by the difficulties in calculating spin-forbidden lines theoretically, uncertainties in the semi-empirical calculations and by the fact that weaker lines have larger uncertainties than strong lines in the experiments. In Fig. ~\ref{fig:comp2b} and Fig. ~\ref{fig:comp2a}, we compared our results derived from the experimental lifetimes with the previous experimental log$(gf)_{PRK}$ values of \citet{Parkinson} and log$(gf)_{LW}$ values of \citet{Lawler} in the optical region. Comparisons with experimental values show very good agreement, especially with \citet{Lawler}. This supports that our results are accurate in the infrared region. The small uncertainties for the oscillator strengths presented in this work allow for improved stellar abundances.

\begin{acknowledgements}
We acknowledge the grant no $621-2011-4206$ from the Swedish Research Council (VR) and support from The Gyllenstierna Krapperup's Foundation. The infrared FTS at the Edl{\'e}n laboratory is made available through a grant from the Knut and Alice Wallenberg Foundation. HN acknowledges the funding from the Swedish Research Council through the Linnaeus grant to the Lund Laser Centre.
\end{acknowledgements}

\bibliography{bibli} 
\bibliographystyle{aa}

\end{document}